\documentclass[letterpaper,twoside,english]{article}
\usepackage[T1]{fontenc}
\usepackage{amsmath}
\usepackage{amssymb}
\IfFileExists{url.sty}{\usepackage{url}}
                      {\newcommand{\url}{\texttt}}

\makeatletter
\AtBeginDocument{

}

\usepackage{babel}
\makeatother
\begin{document}

\newcommand{\set}[2]{\left\{  #1\, |\, #2\right\}  }

\newcommand{\cyc}[1]{\mathbb{Q}\left[\zeta_{#1}\right]}
 
\newcommand{\triv}{1}

\newcommand{\Mod}[3]{#1\equiv#2\, \left(\mathrm{mod}\, \, #3\right)}

\newcommand{\X}[2]{Y\left(#1,#2\right)}

\newcommand{\Y}[3]{\Upsilon_{#3}\left(#1,#2\right)}

\newcommand{\trans}[2]{\mathcal{W}\left[#1,#2\right]}

\newcommand{\FA}[1]{\vartheta}

\newcommand{\FB}[1]{\sqrt{G}}

\newcommand{\FC}[1]{\mathbb{Z}_{2}}

\newcommand{\FD}[1]{\Pi\left(#1\right)}

\newcommand{\FE}[2]{\mathrm{rad}}

\newcommand{\FF}[2]{\mathrm{Fix}_{#1}\left(#2\right)}

\newcommand{\irr}{\mathrm{Irr}\left(G,\FA{}\right)}

\newcommand{\mult}[2]{\mathrm{mult}_{\mathcal{X}}\left(H,#2\right)}

\newcommand{\pe}[1]{#1^{\perp}}

\newcommand{\perm}[1]{\Pi\left(#1\right)}

\newcommand{\thet}{\sqrt{\FA{}}}

\title{Simple current symmetries in RCFT}

\author{P. Bantay\\
Institute for Theoretical Physics, Eotvos University Budapest}

\maketitle
\begin{abstract}
The question ''Which abelian permutation groups arise as group of
simple currents in Rational Conformal Field Theory?'' is investigated
using the formalism of weighted permutation actions. After a review
of the relevant properties of simple current symmetries, the general
theory of WPA-s and admissibility conditions are described, and classification
results are illustrated by a couple of examples.
\end{abstract}

\section{Introduction\label{sec:Introduction}}

The global structure of a Rational Conformal Field Theory is described
by a Modular Tensor Category \cite{Turaev,MTC}, i.e. a solution of
the Moore-Seiberg polynomial equations \cite{MS} and the associated
modular representation. Besides the behavior of the genus one characters
under the action of the modular group \cite{Cardy}, the Modular Tensor
Category does also determine the fusion rules of the primary fields
\cite{Verlinde}, the modular invariant partition functions \cite{modinv},
the conformal boundary conditions \cite{boundary,boundary2}, and
many other important characteristics of the theory.

Recall that the modular representation associated to a RCFT is determined
by a pair of unitary matrices $S$ and $T$ representing the modular
transformations $\tau\mapsto\frac{-1}{\tau}$ and $\tau\mapsto\tau+1$
respectively (with rows and columns indexed by the primary fields),
such that \cite{MS,Verlinde}

\begin{enumerate}
\item $T$ is diagonal of finite order, and $S$ is symmetric;
\item $S^{2}$ commutes with $T$, and equals the permutation matrix associated
to charge conjugation;
\item the modular relation \begin{equation}
STS=T^{-1}ST^{-1}\label{modrel}\end{equation}
holds.
\end{enumerate}
The eigenvalues of the matrix $T$ are determined by the conformal
weights $\Delta_{p}$ of the primaries and the central charge $c$,
more precisely \begin{equation}
T_{p}^{q}=\omega\left(p\right)\delta_{p}^{q}\,\,\label{tpq}\end{equation}
with \begin{equation}
\omega\left(p\right)=\exp\left(2\pi\imath\left(\Delta_{p}-\frac{c}{24}\right)\right)\,\,,\label{omdef}\end{equation}
 while the matrix elements of $S$ determine the fusion rules $N_{pqr}$
of the primaries by Verlinde's formula \cite{Verlinde}\begin{equation}
N_{pqr}=\sum_{s}\frac{S_{ps}S_{qs}S_{rs}}{S_{0s}}\,\,,\label{Ver}\end{equation}
where $0$ labels the vacuum.

Symmetries of the modular representation have been studied extensively,
mostly because of their intimate ties with the classification of modular
invariant partition functions. A prominent role is played by simple
current symmetries, which are permutations $\alpha$ of the primary
fields for which the associated permutation matrix \begin{equation}
\left[\FD{\alpha}\right]_{p}^{q}=\delta_{p}^{\alpha q}\,\,\,\label{eq:permmat}\end{equation}
is diagonalized by $S$. The terminology comes from the relation of
such permutations to simple currents, which are defined traditionally
as primary fields $J$ whose fusion product with their charge conjugate
equals the vacuum \cite{SC1,SC2,SC3,SC4}, a condition that may be
shown to be equivalent to $S_{0J}=S_{00}$. From this definition follows,
using standard properties of the modular representation and Verlinde's
formula Eq.(\ref{Ver}), that the fusion product of two simple currents
is again a simple current, i.e. they form an abelian group, and that
the fusion matrix $N_{J}$ associated to a simple current $J$ is
a permutation matrix. In other words, to any simple current $J$ is
associated a permutation $\alpha_{J}$ of the primary fields such
that $N_{J}=\Pi\left(\alpha_{J}\right)$. According to Verlinde's
theorem \cite{Verlinde} fusion matrices are diagonalized by $S$,
consequently $\alpha_{J}$ is a simple current symmetry for each $J$.
Conversely, $\alpha0$ is a simple current for each simple current
symmetry $\alpha$, and this one-to-one correspondence between simple
currents and simple current symmetries shows that the two notions
are in fact equivalent. We'll use the definition based on permutations
because it is more adequate for the kind of questions we intend to
discuss. 

Simple currents play a distinguished role in Conformal Field Theory
for several reasons. Just to cite a few, they provide for most modular
invariant partition functions \cite{SCinv,SCinv2}, they are essential
for the proper treatment of the GSO projection in Superstring Theory
\cite{GSO,GSO1,GSO2}, they can be used to construct new non-trivial
modular tensor categories \cite{SCEXT}, and they determine to a great
extent the structure of the modular representation \cite{projker}.
All these applications need a thorough and detailed understanding
of their properties.

While there is an extensive literature on simple currents \cite{SC1,SC2,SC3,SC4,marty,Picard,gannon},
the following question still waits for an answer.

\begin{description}
\item [Question1:]\label{Q1}Which finite abelian permutation groups are
group of simple current symmetries of a RCFT? 
\end{description}
That any (finite) abelian group arises this way is easy to see, but
it is far from true that any abelian permutation group does. For example,
we'll see that the degree of the group of simple current symmetries
is bounded from below by its order, restricting severely the possible
permutation groups. 

A simple minded approach to answering Question1 would be to classify
all RCFT-s (or the corresponding modular tensor categories), and to
compute the group of simple current symmetries for each of them: sorting
the resulting list would yield the answer. Unfortunately, the enumeration
of all RCFT-s is still beyond reach, so we need some alternative strategy.
The one we'll adopt here is to look for properties that are characteristic
of the permutation action of the group of simple current symmetries,
then, for a prescribed abelian group, to classify all permutation
actions with these properties. As we'll see, these characteristic
properties are related to the extra structure of simple current symmetries,
formalized in the notion of an admissible weighted permutation action.
In this way we'll get an answer to our problem, which is only partial
because not all admissible WPA-s correspond to simple current symmetries
of RCFT-s. Nevertheless, this partial answer will already allow us
to draw surprising conclusions about RCFT-s with a prescribed group
of simple current symmetries, e.g. that a theory which has a $\mathbb{Z}_{3}\times\mathbb{Z}_{3}$
group of integer spin simple currents should have at least 35 primary
fields (Example 4 of Section \ref{sec:Examples}), a result that is
not completely obvious. But first of all, we have to take a closer
look at simple current symmetries.

\section{Simple current symmetries\label{sec:Simple-current-symmetries}}

As explained in Section \ref{sec:Introduction}, a simple current
symmetry is a permutation $\alpha$ of the primary fields such that
$S^{-1}\FD{\alpha}S$ is a diagonal matrix, where $\FD{\alpha}$ denotes
the permutation matrix associated to $\alpha$. A straightforward
argument shows that simple current symmetries form an abelian group
$G$ (commutativity being a consequence of the commutativity of diagonal
matrices). The importance of simple current symmetries is related
to the following result, somewhat reminiscent of Schur's lemma: any
monomial matrix (i.e. a matrix that has exactly one nonzero entry
in each row and column) diagonalized by $S$ is necessarily proportional
to $\FD{\alpha}$ for some $\alpha\in G$.

The basic properties of simple current symmetries are neatly expressed
in terms of the monomial matrices \begin{equation}
\X{\alpha}{\beta}=\FD{\alpha}S\FD{\beta}S^{-1}\,\,.\label{eq:Xidef}\end{equation}
A clever application of the modular relation Eq.(\ref{modrel}), together
with Eq.(\ref{tpq}) and the symmetry of $S$, leads to the formula\begin{equation}
\X{\alpha}{\beta}_{p}^{q}=\FA{}\left(\beta\right)\frac{\omega\left(q\right)}{\omega\left(\beta q\right)}\delta_{p}^{\alpha q}\,\,\label{Xiom}\end{equation}
for the matrix elements, where\begin{equation}
\FA{}\left(\alpha\right)=\exp\left(2\pi\imath\Delta_{\alpha0}\right)\,\,.\label{eq:thdef}\end{equation}
Inserting Eq.(\ref{Xiom}) into the obvious equality $\X{1}{\alpha}\X{1}{\beta}=\X{1}{\alpha\beta}$
gives the fundamental relation\begin{equation}
\frac{\omega\left(\alpha p\right)\omega\left(\beta p\right)}{\omega\left(p\right)\omega\left(\alpha\beta p\right)}=\frac{\FA{}\left(\alpha\right)\FA{}\left(\beta\right)}{\FA{}\left(\alpha\beta\right)}\,\,,\label{eq:omeq}\end{equation}
which holds for all primaries $p$ and all $\alpha,\beta\in G$, where
the point is that the lhs. of Eq.(\ref{eq:omeq}) does not depend
on $p$.

Eq.(\ref{eq:omeq}) is the cornerstone of the whole theory. In particular,
from Eqs.(\ref{Xiom}) and (\ref{eq:omeq}) one derives the multiplication
rule\begin{equation}
\X{\alpha_{1}}{\beta_{1}}\X{\alpha_{2}}{\beta_{2}}=\frac{\FA{}\left(\alpha_{2}\right)\FA{}\left(\beta_{1}\right)}{\FA{}\left(\alpha_{2}\beta_{1}\right)}\X{\alpha_{1}\alpha_{2}}{\beta_{1}\beta_{2}}\,\,.\label{Ximult}\end{equation}
In other words the matrices $\X{\alpha}{\beta}$ form a projective
representation of the abelian group $G\times G$, with a suitable
2-cocycle determined by $\FA{}$. In terms of $\FA{}$ the 2-cocycle
condition reads 

\begin{equation}
\FA{}\left(\alpha\beta\right)\FA{}\left(\beta\gamma\right)\FA{}\left(\gamma\alpha\right)=\FA{}\left(\alpha\right)\FA{}\left(\beta\right)\FA{}\left(\gamma\right)\FA{}\left(\alpha\beta\gamma\right)\:,\label{eq:thmult}\end{equation}
 i.e. $\FA{}$ is a quadratic function on $G$ written multiplicatively!
The 2-cocycle of Eq.(\ref{Ximult}) is trivial if and only if $\FA{}$
is a character of $G$, i.e. $\FA{}\left( \alpha\beta\right)=\FA{}\left(\alpha\right)\FA{}\left(\beta\right)$.
One may also show that in a unitary RCFT\begin{equation}
\FA{}\left(\alpha^{n}\right)=\FA{}\left(\alpha\right)^{n^{2}}\label{homog1}\end{equation}
holds for all integers $n$ and $\alpha\in G$, i.e. $\FA{}$ is actually
a quadratic form on $G$, a well known result of the theory of simple
currents \cite{SC1,SC3}.

Let's now consider the commutation rule of $\X{\alpha}{\beta}$ with
modular representation matrices. If $M$ represents the $SL\left(2,\mathbb{Z}\right)$
element$\left(\begin{array}{cc}
a & b\\
c & d\end{array}\right)$, then \begin{equation}
M^{-1}\X{\alpha}{\beta}M=\frac{\FA{}\left(\alpha\right)^{b\left(c-a\right)}\FA{}\left(\beta\right)^{c\left(b-d\right)}}{\FA{}\left(\alpha\beta\right)^{bc}}\X{\alpha^{a}\beta^{c}}{\alpha^{b}\beta^{d}}\,\,.\label{Ximod}\end{equation}
In particular, \begin{equation}
S^{-1}\X{\alpha}{\beta}S=\frac{\FA{}\left(\alpha\beta\right)}{\FA{}\left(\alpha\right)\FA{}\left(\beta\right)}\X{\beta}{\alpha^{-1}}\label{eq:XiS}\end{equation}
and \begin{equation}
T^{-1}\X{\alpha}{\beta}T=\frac{1}{\FA{}\left(\alpha\right)}\X{\alpha}{\alpha\beta}\,\,.\label{eq:XiT}\end{equation}

Eq.(\ref{Ximod}) is the basis of the application of simple currents
to the construction of modular invariant partition functions \cite{SCinv,SCinv2}:
if $H<G$ is a subgroup such that $\FA{}\left(\alpha\right)=1$ for
all $\alpha\in H$ (i.e. $H$ corresponds to integer spin simple currents),
then the matrix \begin{equation}
Z_{H}=\frac{1}{\left|H\right|}\sum_{\alpha,\beta\in H}\X{\alpha}{\beta}\,\,\label{eq:Scinv}\end{equation}
commutes with all modular matrices according to Eq.(\ref{Xiom}),
and may be shown to have non-negative integer entries, thus providing
a modular invariant, the so-called simple current modular invariant
associated to the subgroup $H$. There is a further possibility to
introduce into Eq.(\ref{eq:Scinv}) so-called ''discrete torsion''
coefficients \cite{SCinv2}.

An important feature of the matrices $\X{\alpha}{\beta}$, which goes
beyond the usual properties of the modular representation, follows
from considerations involving the mapping class group action on the
space of genus one 1-point holomorphic blocks, more precisely from
the ''curious relation'' Eq.(33) of \cite{SCext}. According to
this result, the trace of $\X{\alpha}{\beta}$ may be expressed as\begin{equation}
\mathrm{Tr\,\,}\X{\alpha}{\beta}=\sum_{p}\phi_{p}\left(\alpha,\beta\right)\,\,,\label{eq:UpsFi}\end{equation}
where the quantity $\phi_{p}\left(\alpha,\beta\right)$, the so-called
''commutator cocycle'', satisfies \begin{equation}
\left|\phi_{p}\left(\alpha,\beta\right)\right|=\begin{cases}
1 & \mathrm{if}\,\, p\,\,\mathrm{is\,\, fixed\,\, by\,\, both}\,\,\alpha\,\,\mathrm{and}\,\,\beta,\\
0 & \mathrm{otherwise}.\end{cases}\label{eq:FiBound}\end{equation}

Finally, one may show that no simple current symmetry besides the
identity may fix the vacuum. The importance of this remark will be
clarified in Section \ref{sec:Admissible-WPAs}, but it does already
imply that there are at least as many primary fields as simple currents,
i.e. the degree of the permutation group $G$ cannot be less than
its order.

\section{Weighted permutation actions\label{sec:Weighted-permutation-actions} }

By a quadratic group we'll mean a pair $\left(G,\FA{}\right)$, with
$\FA{}$ a quadratic function on the finite abelian group $G$, i.e.
a complex valued function $\FA{}:G\rightarrow\mathbb{C}^{*}$ such
that Eq.(\ref{eq:thmult}) is satisfied. Its radical is the quadratic
group \begin{equation}
\FE{}{}\left(G,\FA{}\right)=\left(\FB{},\thet\right)\,\,,\label{eq:raddef}\end{equation}
where $\FB{}=\set{\alpha\in G}{\FA{}\left(\alpha\beta\right)=\FA{}\left(\alpha\right)\FA{}\left(\beta\right)\,\,\,\forall\beta\in G}$
and $\thet$ is the restriction of $\FA{}$ to $\FB{}$. Note that
$\thet$, besides being a quadratic function, is also a character
of $\FB{}$, i.e. it is a homomorphism from $\FB{}$ into the complex
numbers. A quadratic group is called non-degenerate if its radical
is trivial, while it is called completely degenerate if it equals
its own radical (in which case the quadratic function $\FA{}$ is
a character of $G$). While usually the nondegenerate case is the
interesting one, we'll see that in the present context it's just the
opposite: all the interest lies in the completely degenerate case. 

By a weighted permutation action (WPA for short) of a quadratic group
$\left(G,\FA{}\right)$ we shall mean a pair $\mathcal{X}=\left(X,\omega\right)$,
where $X$ (the support of $\mathcal{X}$) is a finite set permuted
by the group $G$, while $\omega$ (the weight function of $\mathcal{X}$)
is a nowhere vanishing complex function on $X$ which satisfies Eq.(\ref{eq:omeq}).
The degree of the WPA $\left(X,\omega\right)$ is the cardinality
of $X$. According to the results of Section \ref{sec:Simple-current-symmetries},
the group $G$ of simple current symmetries of a RCFT, together with
the quadratic function $\FA{}$ defined by Eq.(\ref{eq:thdef}), form
a quadratic group $\left(G,\FA{}\right)$, and the pair $\left(X,\omega\right)$
is a weighted permutation action of this quadratic group, where $X$
denotes the set of primary fields and $\omega$ is defined by Eq.(\ref{omdef}).
We shall call this WPA the simple current WPA associated to the RCFT.

Clearly, for a given WPA $\mathcal{X}=\left(X,\omega\right)$ of the
quadratic group $\left(G,\FA{}\right)$ the labeling of the elements
of $X$ is immaterial, leading to an obvious notion of equivalence
for WPA-s, with an additional freedom: Eq.(\ref{eq:omeq}) is invariant
under a rescaling of $\omega$, therefore two WPA-s whose weight functions
differ by a factor which is locally constant on the orbits of $G$
should be considered equivalent too. Our goal in this Section is to
understand the classification of WPA-s up to equivalence.

First of all, let's see an important example of a WPA. For a given
quadratic group $\left(G,\FA{}\right)$ let's consider the WPA whose
support is the set of elements of $G$ (on which $G$ acts by left
translations), and whose weight function equals $\FA{}$. That this
is indeed a WPA, which we'll call the regular WPA, follows from Eq.(\ref{eq:thmult}).
This WPA is transitive, which means that its support consist of just
one $G$-orbit, and it will play an especially important role later.

Given two WPA-s $\mathcal{X}_{1}=\left(X_{1},\omega_{1}\right)$ and
$\mathcal{X}_{2}=\left(X_{2},\omega_{2}\right)$ with disjoint supports
(i.e. such that $X_{1}\cap X_{2}=\emptyset$), their sum is the WPA
denoted $\mathcal{X}_{1}\oplus\mathcal{X}_{2}$, with support $X_{1}\cup X_{2}$
and weight function\begin{equation}
\omega_{\mathcal{X}}\left(p\right)=\begin{cases}
\omega_{1}(p) & \mathrm{if}\,\, p\in X_{1}\,\,,\\
\omega_{2}(p) & \mathrm{if}\,\, p\in X_{2}\,\,.\end{cases}\label{eq:wpasum}\end{equation}
It may be shown that one gets in this way a commutative and associative
binary operation on the set of equivalence classes of WPA-s. There
is also an obvious notion of product of WPA-s, but it won't play any
role in what follows.

Transitive WPA-s, i.e. those WPA-s whose support consist of a single
$G$-orbit, are fundamental because any WPA may be decomposed uniquely
(up to reordering) into a sum of transitive ones. This result reduces
the classification of (equivalence classes of) WPA-s to the problem
of determining the transitives.

All transitive WPA-s may be obtained (up to equivalence) as follows:
let $\xi$ be a character of $\sqrt{G}$, and $H$ a subgroup of $\ker\left(\xi/\sqrt{\FA{}}\right)$
(the latter makes sense since $\sqrt{\FA{}}$ is a character of $\FB{}$).
Consider the coset space $X=G/H$ on which $G$ acts by left translations,
and the function $\omega\left(\alpha H\right)=\FA{}\left(\alpha\right)/\xi^{*}\left(\alpha\right)$,
where $\xi^{*}$ is any character of $G$ whose restriction to $\FB{}$
equals $\xi$. One may show that this $\omega$ is well-defined on
$G/H$, and the pair $\left(X,\omega\right)$ is a transitive WPA,
whose equivalence class $\trans{H}{\xi}$ is fully determined by $H$
and $\xi$, and any transitive WPA is equivalent to one of these coset
WPA-s. The regular WPA is just $\trans{\triv}{\xi_{0}}$, where $\xi_{0}$
denotes the trivial character, and $\triv$ is the trivial subgroup
of $G$. 

There is a result that simplifies greatly the classification of WPA-s.
It states that there is a one-to-one correspondence between the WPA-s
of a quadratic group and those of its radical, which respects sums,
and under which transitive WPA-s correspond to transitive ones! In
particular, a non-degenerate quadratic group has just one transitive
WPA (the regular one), all other WPA-s being multiples of it, showing
that, as far as WPA-s are concerned, the interest does indeed lie
in the completely degenerate case. 

In the same way as one can associate a linear representation to any
permutation action of a group, one can associate a monomial representation
to a WPA $\left(X,\omega\right)$ by introducing matrices\begin{equation}
\X{\alpha}{\beta}_{p}^{q}=\FA{}\left(\beta\right)\frac{\omega\left(q\right)}{\omega\left(\beta q\right)}\delta_{p}^{\alpha q}\,\,,\label{Xiom2}\end{equation}
whose rows and columns are labeled by the elements of $X$. These
matrices form a projective representation of $G\times G$, i.e.\begin{equation}
\X{\alpha_{1}}{\beta_{1}}\X{\alpha_{2}}{\beta_{2}}=\frac{\FA{}\left(\alpha_{2}\right)\FA{}\left(\beta_{1}\right)}{\FA{}\left(\alpha_{2}\beta_{1}\right)}\X{\alpha_{1}\alpha_{2}}{\beta_{1}\beta_{2}}\,\,.\label{Ximult2}\end{equation}
 Moreover, denoting by $A^{t}$ the transpose of a matrix $A$, it
follows from Eqs.(\ref{Xiom2}) and (\ref{eq:omeq}) that \begin{equation}
\X{\alpha}{\beta}^{t}=\frac{\FA{}\left(\alpha\beta\right)}{\FA{}\left(\alpha\right)\FA{}\left(\beta\right)}\X{\alpha^{-1}}{\beta}\,\,.\label{eq:Xitranspose}\end{equation}
The quantity \begin{equation}
\Y{\alpha}{\beta}{\mathcal{X}}=\FA{}\left(\beta\right)\sum_{p\in\FF{}{\alpha}}\frac{\omega\left(p\right)}{\omega\left(\beta p\right)}=\mathrm{Tr\,\,}\X{\alpha}{\beta}\,\,,\label{eq:Ydef2}\end{equation}
where $\FF{}{\alpha}$ denotes the set of fixed points of the permutation
$\alpha$, will play an important role in Section \ref{sec:Admissible-WPAs}.
It is obviously additive, i.e.\begin{equation}
\Y{\alpha}{\beta}{\mathcal{X}}=\Y{\alpha}{\beta}{\mathcal{X}_{1}}+\Y{\alpha}{\beta}{\mathcal{X}_{2}}\,\,\label{eq:upsadd}\end{equation}
if $\mathcal{X}=\mathcal{X}_{1}\oplus\mathcal{X}_{2}$, and for the
transitive WPA $\trans{H}{\xi}$ (see Section \ref{sec:Weighted-permutation-actions})
one has\begin{equation}
\Y{\alpha}{\beta}{}=\begin{cases}
\xi\left(\beta\right)\left[G:H\right] & \mathrm{if}\,\,\alpha\in H\,\,\mathrm{and}\,\,\beta\in\FB{},\\
0 & \mathrm{otherwise}.\end{cases}\label{eq:cosups}\end{equation}
Note that $\Y{\alpha}{\beta}{}=0$ unless $\alpha,\beta\in\FB{}$,
consequently we may consider $\Upsilon$ as a function defined on
$\FB{}\times\FB{}$.

\section{Admissible WPAs\label{sec:Admissible-WPAs}}

As we have already discussed in Section \ref{sec:Weighted-permutation-actions},
to each RCFT corresponds a WPA $\left(X,\omega\right)$ of the quadratic
group $\left(G,\FA{}\right)$, where $X$ is the set of primary fields,
$\omega$ is defined by Eq.(\ref{omdef}), $G$ is the group of simple
current symmetries, and $\FA{}$ the quadratic function defined by
Eq.(\ref{eq:thdef}). In view of this result, we may reformulate Question1
from Section \ref{sec:Introduction} as follows. 

\begin{description}
\item [Question2:]\label{Q2}For a given a quadratic group, which of its
WPA-s arise as simple current WPA-s of some RCFT?
\end{description}
It is clear that an arbitrary WPA won't do the job. First, because
the vacuum $0$ has trivial stabilizer (as noticed at the end of Section
\ref{sec:Simple-current-symmetries}), and the weight function $\omega$
restricted to the $G$-orbit of $0$ equals, up to an irrelevant constant
of proportionality, the quadratic form $\FA{}$ according to Eqs.(\ref{omdef})
and (\ref{eq:thdef}), it follows that the simple current WPA of a
unitary RCFT should contain at least one copy of the regular WPA.
But this is not the end of the story.

An analysis of the properties of simple current WPA-s leads to the
notion of admissible WPA, by which we mean a WPA $\mathcal{X}$ that
satisfies the following criteria:

\begin{enumerate}
\item Galois invariance: \begin{equation}
\mult{\mathcal{X}}{\xi^{k}}=\mult{\mathcal{X}}{\xi}\label{eq:galinv}\end{equation}
for any integer $k$ coprime to the exponent of $G$, where $\mult{\mathcal{X}}{\xi}$
denotes the multiplicity of the transitive WPA $\trans{H}{\xi}$ in
$\mathcal{X}$.
\item Reciprocity relation: \begin{equation}
\Y{\beta}{\alpha}{}=\Y{\alpha}{\beta}{}\,\,.\label{eq:reciprocity}\end{equation}

\item Fixed point bound: \begin{equation}
\left|\Y{\alpha}{\beta}{}\right|\leq\left|\FF{}{\alpha}\cap\FF{}{\beta}\right|\label{eq:ybound}\end{equation}
for all $\alpha,\beta\in G$, i.e. the absolute value of $\Y{\alpha}{\beta}{}$
is bounded from above by the number of common fixed points of $\alpha$
and $\beta$.
\end{enumerate}
Let's explain the origin of these conditions! Eq.(\ref{eq:galinv})
follows ultimately from Eq.(\ref{Ximod}), for one can show using
the theory of the Galois action \cite{Galois,Galois2}, that the matrix
$S^{-1}T^{l}ST^{k}ST^{l}$ is monomial for integers $l$ and $k$
such that $\Mod{lk}{1}{N}$, where $N$ denotes the order of the matrix
$T$ \cite{Modker}. Because the exponent of the group $G$ of simple
currents divides $N$, Eq.(\ref{Ximod}) applied to such matrices
$S^{-1}T^{l}ST^{k}ST^{l}$, taking into account their monomiality,
leads to Eq.(\ref{eq:galinv}) for the simple current WPA of a RCFT. 

As to Eq.(\ref{eq:reciprocity}), it comes from combining Eqs.(\ref{eq:XiS})
and (\ref{eq:Xitranspose}), which give \begin{equation}
S^{-1}\X{\alpha}{\beta}^{t}S=\X{\beta}{\alpha}\,\,.\label{Xirecip}\end{equation}
Taking the trace of both sides and recalling Eq.(\ref{eq:Ydef2})
gives at once the reciprocity relation Eq.(\ref{eq:reciprocity}).
Finally, the fixed point bound follows from Eqs.(\ref{eq:UpsFi}),
(\ref{eq:FiBound}) and the triangle inequality.

According to the above terminology, the simple current WPA of a unitary
RCFT is an admissible WPA which contains the regular WPA with positive
multiplicity. Note that the above admissibility conditions are necessary,
but by no means sufficient! 

One might wonder how restrictive is the above concept of admissibility,
and on the possible dependencies between Eqs.(\ref{eq:galinv}), (\ref{eq:reciprocity})
and (\ref{eq:ybound}). One may show that they are actually independent
by exhibiting relatively simple examples which satisfy two of them,
but not the third. As to the strength of these restrictions, let's
just mention the following result: for any WPA that satisfies the
reciprocity relation Eq.(\ref{eq:reciprocity}), the values of $\Y{\alpha}{\beta}{}$
are always integers! The examples from Section \ref{sec:Examples}
further illustrate the special nature of admissible WPA-s. In principle
one could develop more restrictive criteria for WPA-s corresponding
to simple current symmetries of a RCFT, but the above notion of admissibility
has a distinctive feature: it puts linear restrictions (equalities
and inequalities) on the multiplicities of the transitive constituents,
which means that the sum of two admissible WPA-s is again admissible.

Let's turn now to the classification of admissible WPA-s. For this
we'll need the notion of irreducibility: a WPA is irreducible if it
is admissible and cannot be decomposed into a non-trivial sum of two
admissible WPA-s. One can show that there is only a finite number
of irreducibles, which follows from the fact that the admissibility
conditions translate into a set of homogeneous linear equalities and
inequalities on the multiplicities of the transitive constituents,
and these define a rational cone which, according to the celebrated
theorem of Minkowski and Weyl \cite{polytopes}, is always generated
by a finite number of extreme rays .

Consequently, it is in principle possible to determine for each quadratic
group the set of irreducible WPA-s, and any admissible WPA will arise
as a sum of these irreducibles. But it should be noticed that, although
an irreducible decomposition always exists for an admissible WPA,
this decomposition is by no means unique, for there may exist non-trivial
relations among the irreducibles (as in Example 4 of the next Section).

An important simplification results from the one-to-one correspondence
(discussed in Section \ref{sec:Weighted-permutation-actions}) between
the WPA-s of a quadratic group and those of its radical, for irreducible
WPA-s correspond under it to irreducible ones. Thus it is enough to
classify the irreducibles WPA-s of the radical. In particular, for
a nondegenerate quadratic group the regular WPA, which is the only
transitive, is also the only irreducible, which corresponds to the
situation where all primaries are simple currents. Conversely, the
regular WPA cannot be admissible unless the quadratic group is nondegenerate.

\section{\label{sec:Examples}Examples}

As indicated in the main text, there is a one-to-one correspondence
between the WPA-s of a quadratic group and the WPA-s of its radical,
under which transitive (resp. irreducible) WPA-s correspond to transitive
(resp. irreducible) ones. Thus it is enough to treat the completely
degenerate case, when the quadratic function is a character. As usual,
$\mathbb{Z}_{n}$ is the cyclic group of order $n$, and we'll denote
by $\xi_{0}$ the trivial character of any group, while $\xi_{1}$
stands for the non-trivial character of $\mathbb{Z}_{2}$. 

\begin{enumerate}
\item $\mathcal{G}=\left(\FC{},\xi_{0}\right)$.\\
This quadratic group has three transitive WPA-s (see Section \ref{sec:Weighted-permutation-actions}
for notations): $F=\trans{\FC{}}{\xi_{0}}$ of degree $1$ (the fixed
point), $R=\trans{\triv}{\xi_{0}}$ of degree $2$ (the regular),
and $W=\trans{\triv}{\xi_{1}}$ of degree $2$. The Galois action
is trivial because $G$ has exponent $2$. A WPA \[
n_{F}F\oplus n_{R}R\oplus n_{W}W\,\,,\]
is admissible if and only if the following conditions are satisfied:
\begin{eqnarray*}
n_{R} & = & n_{W}\,\,,\\
n_{F},n_{R} & \geq & 0\,\,.\end{eqnarray*}
It does follow that there are two irreducible WPA-s: $F$ of degree
1 and $R\oplus W$ of degree 4, and there are no relations between
them. Any admissible WPA that contains the regular one does also contain
the WPA $W$, in particular its degree is at least $4$.
\item $\mathcal{G}=\left(\FC{},\xi_{1}\right)$.\\
Again we have three transitive WPA-s, $F=\trans{\FC{}}{\xi_{1}}$,
$R=\trans{\triv}{\xi_{0}}$ and $W=\trans{\triv}{\xi_{1}}$, out of
which $R$ is the regular of degree $2$, and $F$ is the fixed-point
WPA of degree $1$. The admissibility conditions for $n_{F}F\oplus n_{R}R\oplus n_{W}W$
read\begin{eqnarray*}
n_{R} & = & n_{F}+n_{W}\,\,,\\
n_{F},n_{W} & \geq & 0\,\,.\end{eqnarray*}
Consequently, there are two irreducible WPA-s: $F\oplus R$ of degree
$3$ and $R\oplus W$ of degree $4$, and there is no relation between
them. In this case both irreducibles contain the regular WPA $R$.
The irreducible $F\oplus R$ is realized in the Ising model.
\item $\mathcal{G}=\left(\FC{}\times\FC{},\xi_{0}\right)$.\\
In this case there are 11 transitive WPA-s. Solving the conditions
for admissibility, we get a total of 6 irreducible WPA-s, out of which
three have degree $4$, and the remaining three have degrees $1,11$
and $16$ respectively. There are no relations between the irreducibles.
The only irreducibles that contain the regular WPA are those of degrees
11 and 16. The irreducible of degree 11 is realized in a suitable
Ashkin-Teller model.
\item $\mathcal{G}=\left(\mathbb{Z}_{3}\times\mathbb{Z}_{3},\xi_{0}\right)$.\\
There are 22 transitive WPA-s belonging to 14 Galois orbits, but only
8 irreducible ones. Of the irreducible WPA-s four have degree 9, while
the remaining have respective degrees 1, 35, 69 and 81. The only irreducible
WPA-s containing the regular one are those of degrees 35, 69 and 81.
There is one non-trivial relation between the irreducibles, which
reads $\underline{35}\oplus\underline{35}=\underline{1}\oplus\underline{69}$
if $\underline{d}$ denotes the unique irreducible of degree $d$.
\end{enumerate}
While the simplest cases with $G=\mathbb{Z}_{2}$ may be dealt with
by hand, the larger cases (large means $\left|G\right|>3$) require
the use of sophisticated software. The needed group theoretic computations
have been performed using GAP \cite{GAP}, while the solution of the
inequalities coming from Eq.(\ref{eq:ybound}) were computed by the
software package cdd+ \cite{CDD}. It should be noted that this last
part of the computations is in general extremely hard. For example,
in the cases $\mathcal{G}=(\mathbb{Z}_{2}\times\mathbb{Z}_{2}\times\mathbb{Z}_{2},\xi_{0})$
and $\mathcal{G}=(\mathbb{Z}_{4}\times\mathbb{Z}_{4},\xi_{0})$ more
than 800 hours of CPU time have not been enough to complete the computation
on a 2.4GHz Pentium4 PC.

\section{\label{sec:Discussion}Discussion}

The main goal of this work was to isolate the notions needed for a
proper treatment of the question: which finite abelian permutation
groups arise as group of simple current symmetries? The analysis of
the properties of simple current symmetries led us to introduce the
notions of weighted permutation actions and admissibility, and to
develop the ensuing theory. While admissibility is a necessary but
not sufficient condition for a WPA to be the simple current WPA of
some RCFT, it has the enormous advantage of being linear in the multiplicities
of the transitive constituents, which greatly simplifies the theory.
Unfortunately, the solution of the linear inequalities coming from
the fixed point bound Eq.(\ref{eq:ybound}) turns out to be very involved
even in simple cases like $G=\mathbb{Z}_{2}\times\mathbb{Z}_{2}\times\mathbb{Z}_{2}$.
But in principle we can determine in finite time the irreducible WPA-s
of any quadratic group, together with the relations satisfied by them,
providing a complete solution of the admissibility conditions and
an answer to Question2. While the notion admissibility is not restrictive
enough, i.e. it may happen that some admissible WPA-s do not arise
from a RCFT, it does nevertheless capture many non-trivial features
of simple current WPA-s, hopefully the most important ones. 

\bigskip{}
\begin{flushleft}\emph{Work supported by grants OTKA T32453, T47041,
T37674 and T43582.}\end{flushleft}

\end{document}